\documentclass[aps,prl,twocolumn,groupedaddress]{revtex4-1}
\usepackage{graphicx}
\usepackage{amssymb}
\usepackage{amsmath}
\usepackage{bm}
\usepackage{epsf}
\usepackage{cancel}

\begin{document}

\title{Thermal elastic-wave attenuation in low-dimensional SiN$_{\rm x}$ bars at low temperatures}



\author{S. Withington,  E. Williams, D. J. Goldie, C. N. Thomas, and M. Schneiderman}
\affiliation{Cavendish Laboratory, University Cambridge, J. J. Thomson Avenue, Cambridge, CB3 OHE}

%\email[Email of Stafford Withington]{stafford@mrao.cam.ac.uk}

\date{\today}

\begin{abstract}
At low temperatures, $<$~200~mK, the thermal flux through low-dimensional amorphous dielectric bars,
$<$~2~$\mu$m wide and 200~nm thick, is transported by a small number of low-order elastic modes. For long bars, $L>$~400~$\mu$m, it is known that the conductance scales as $1/L$, where $L$ is the length, but for short bars, 1~$\mu$m~$<$~$L<$~400~$\mu$m, the length dependence is poorly known. Although it is assumed that the transport must exhibit a diffusive to ballistic transition, the functional form of the transition and the scale size over which the transition occurs have not, to our knowledge, been measured. In this paper, we use ultra-low-noise superconducting Transition Edge Sensors (TESs) to measure the heat flux through a set of SiN$_{\rm x}$ bars to establish the characteristic scale size of the ballistic to diffusive transition. For bars supporting 6 to 7 modes, we measure a thermal elastic-wave attenuation length of 20~$\mu$m. The measurement is important because it sheds light on the scattering processes, which in turn are closely related to the generation of thermal fluctuation noise. Our own interest lies in creating patterned phononic filters for controlling heat flow and thermal noise in ultra-low-noise devices, but the work will be of interest to others trying to isolate devices from their environments, and studying loss mechanisms in micro-mechanical resonators.
\end{abstract}

%\pacs{}

%\keywords{Heat transport, acoustic attenuation length, Transition Edge Sensor}

\maketitle

\section{Introduction}

At low temperatures, $<$~200~mK, low-dimensional dielectric bars transport heat through a small number of elastic modes \cite{ref1,ref2,ref3,ref4,ref5}. The lowest order of these correspond to simple compressional, torsional, and in-plane and out-of-plane flexural waves, and for bars having cross sections of less than about 200 $\times$ 1000~nm, these are the dominant modes present \cite{ref6}. In recent years, low-dimensional dielectric bars have been fabricated in SiN$_{\rm x}$ having lengths $L$ ranging from 400~$\mu$m to 1000~$\mu$m \cite{ref7,ref8}, and it has been found experimentally that, over this range, the low-temperature thermal conductance scales as $1/L$. This length dependence is important because it allows low thermal conductances, $<$ 300~fW K$^{-1}$, to be achieved in nano-engineered components. For example, the sensitivities of far-infrared Transition Edge Sensors (TES) are determined by thermal fluctuation noise in the support legs of the device, and low conductances are needed to achieve ultra-low-noise operation, NEP $<$ 10$^{-18}$ WHz$^{-1/2}$. In addition to measurements on long bars, measurements of conductance and thermal fluctuation noise have been carried out on short bars, 500~nm to 3~$\mu$m, and these show values in precise agreement with ballistic calculations based on the dispersion relations of the low-order elastic waves \cite{ref9}. An outstanding requirement, however, is to understand how transport changes from being fully diffusive to fully ballistic as the length of a bar is reduced, and the characteristic scale length $L_{a}$ over which the transition occurs. The {\em few-mode} ballistic to diffusive transition can only be observed by making measurements on narrow $<$~2~$\mu$m legs having lengths that span the range 5~$\mu$m to 500~$\mu$m.

In this paper we report a set of measurements that reveal the ballistic to diffusive transition, and thereby determine the thermal elastic wave attenuation length $L_{a}$. The uniqueness of our approach is that we determine the phonon-mean-free path (MFP) directly in a few-mode system. We do not infer the MFP indirectly through kinetic arguments based on say heat-capacity measurements on bulk samples.

From a practical perspective, $L_{a}$ is important because it determines how short a bar must be in cases were ballistic transport is needed: for example when fabricating phononic support structures based on say ring resonators, or equivalently Mach-Zhender interferometers. Conversely, it determines how long a bar must be to guarantee a $1/L$ dependence in devices where diffusive transport is needed. More fundamentally, there are still uncertainties about whether the dominant phonon scattering process is elastic or inelastic, and the nature of the transport over the transition region. In particular, $L_{a}$ corresponds to the elastic-wave coherence length, which itself is closely related to the thermal fluctuation noise coupled in and out of a structure. Despite its importance, we are not aware of any experiments that have measured $L_{a}$ directly in a few-mode system, either in SiN$_{\rm x}$ or any other similar amorphous dielectric. Also, we are not aware of any thermal measurements that show the functional form of the diffusive to ballistic transition in low-dimensional structures at low temperatures. The motivation for the work described here was to measure the thermal elastic-wave attenuation length $L_{a}$ of SiN$_{\rm x}$, and to investigate the functional form of the transition region. Although our own interest lies in creating patterned phononic filters for controlling heat flow and noise in low-temperature devices, we believe that the measurement will be of interest to others trying trying to isolate devices from their environments, and studying loss mechanisms in high-Q micro-mechanical resonators \cite{ref10,ref11,ref12}.

\section{Experiment}

We have fabricated a number of Transition Edge Sensors (TESs) having leg lengths ranging from few microns to a few hundred microns: a typical example is shown in the inset of Fig. 1. Each device comprised a superconducting MoAu bilayer and an infrared $\beta$-Ta absorber on a 200~nm SiN$_{\rm x}$ membrane. The absorber is not relevant to the experiment described here, but was included so that the devices could be tested as infrared ( 100-200~$\mu$m ) sensors. The salient dimensions are listed in Table~\ref{table1}. These results supplement previous work \cite{ref9,ref13}, allowing us to build up a complete set of data that spans the range of leg lengths needed.

\begin{table}%[H] add [H] placement to break table across pages
\caption{\label{table1} Device characteristics: $L$, leg length; $W$, leg width; Bars, number of normal-metal bars on bilayer; $K$, power-flow scaling factor; $n$, power-flow temperature dependence; $T_{c}$, critical temperature; $G$, thermal conductance.}
\begin{ruledtabular}
\begin{tabular}{c c c c c c c c}
Device & L & W & Bars & K & n & Tc & G \\
 & ($\mu$m)  & ($\mu$m) &  & (fW K$^{-n}$) & & (mK) & (fW K$^{-1})$ \\ \hline
    1   & 490 & 1.5 & 3 & 115 &  0.99 & 124.9 & 116 \\
    2 & 490 & 2 & 0 & 214.8 & 1.24  & 127.4 & 162 \\
    3 & 490 & 2 & 0 & 660 &  1.61 & 125.9 & 300 \\
    4 & 265 & 1 & 0 & 143 & 0.85  & 126.2 & 166 \\
    5 & 265 & 1 & 6 &  338  & 0.07 & 120.7 & 169\\
\end{tabular}
\end{ruledtabular}
\end{table}

Each TES was biased through two Nb lines that ran along the surfaces of the legs. Electronic heat conduction along the superconducting lines is negligible because the quasiparticle density is exceedingly small at these temperatures. In addition, the Nb leads do not change the elastic modes of the legs because Nb is relatively ductile compared with SiN$_{\rm x}$. Each device was biased using a constant voltage source having a low internal impedance, 1.5~m$\Omega$, which was achieved by integrating a low-value thin-film resistor in an isolated light-tight cavity close to the TES chips. A stray series resistance of typically 2~m$\Omega$ was also measured, and accounted for in the analysis. The devices were read out using SQUIDs as low-noise current to voltage convertors. The whole assembly was contained in a partitioned light-tight box, which was coated on the inside with RF and far-infrared absorber to avoid light leakage placing an additional heat load on the TES island. We have many years of experience fabricating and testing ultra-low-noise TESs, and have detailed models describing all aspects of behaviour \cite{ref14,ref15}. The NEPs of our devices are determined by thermal fluctuation noise in the legs, with different leg conductances giving different sensitivities. The NEPs of the measured devices were all in the range 10$^{-18}$~WHz$^{-1/2}$ to 10$^{-19}$~WHz$^{-1/2}$, depending on the design, which means that small thermal fluxes, in the fW range, could be measured with high precision. The devices were tested in an Adiabatic Demagnetisation Refrigerator (ADR), on a pulse tube cooler, giving a base temperature of around 80~mK. The temperature was stabilised to 200~$\mu$K using a Proportional-Integral-Derivative  (PID) controller to adjust the residual current in the ADR magnet.

\begin{figure}[h]
\noindent \begin{centering}
\includegraphics[trim = 1.0cm 1.0cm 7.0cm 21.0cm,width=7.0cm]{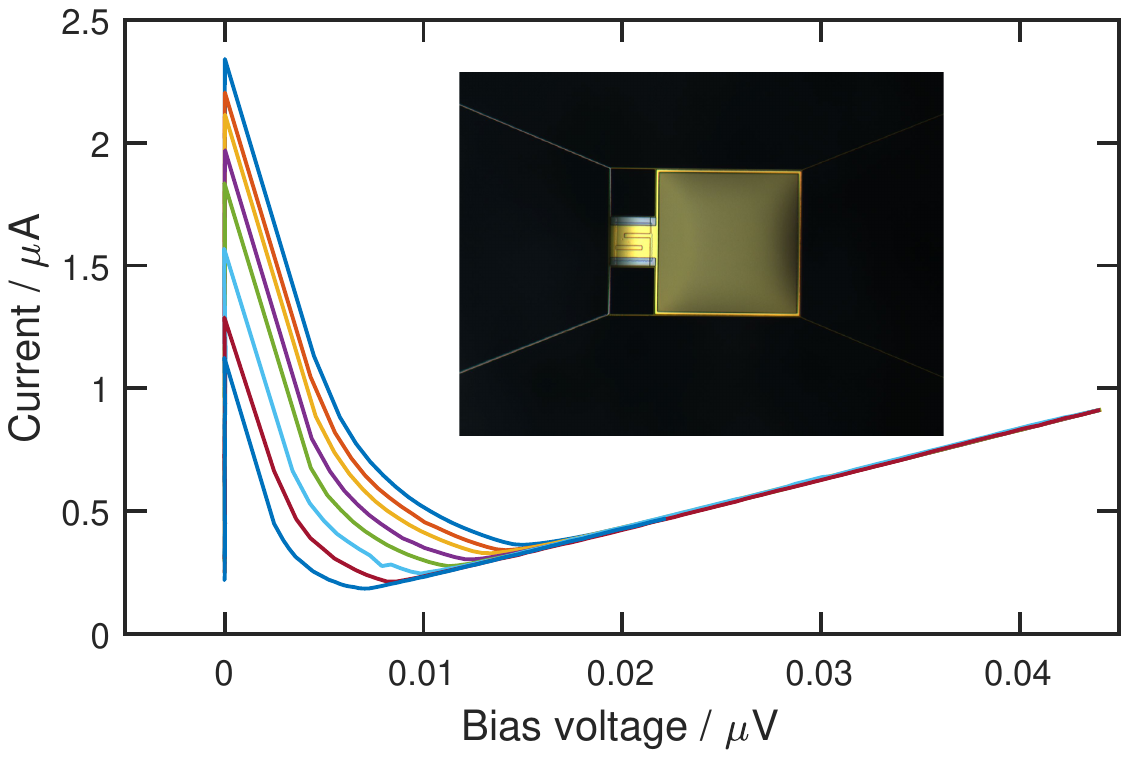}
\par\end{centering}
\caption{Normalised IV curves of one of the MoAu Transition Edge Sensors tested. Different curves correspond to different bath temperatures. The insert shows a photograph of a typical device. The gold-rimmed square is a far-infrared absorber, and not relevant to this experiment. The smaller gold square to the left has side length 160 $\mu$m, and is the MoAu bilayer with addition normal-metal bars.  There are four SiN$_{\rm x}$ legs in total: the two on the left also support Nb wiring. We routinely fabricate long legs having thicknesses of 200~nm
and widths of down to 500~nm.}
\label{figB}
\end{figure}

\begin{figure}[h]
\noindent \begin{centering}
\includegraphics[trim = 1.0cm 1.0cm 7.0cm 21.0cm,width=7.0cm]{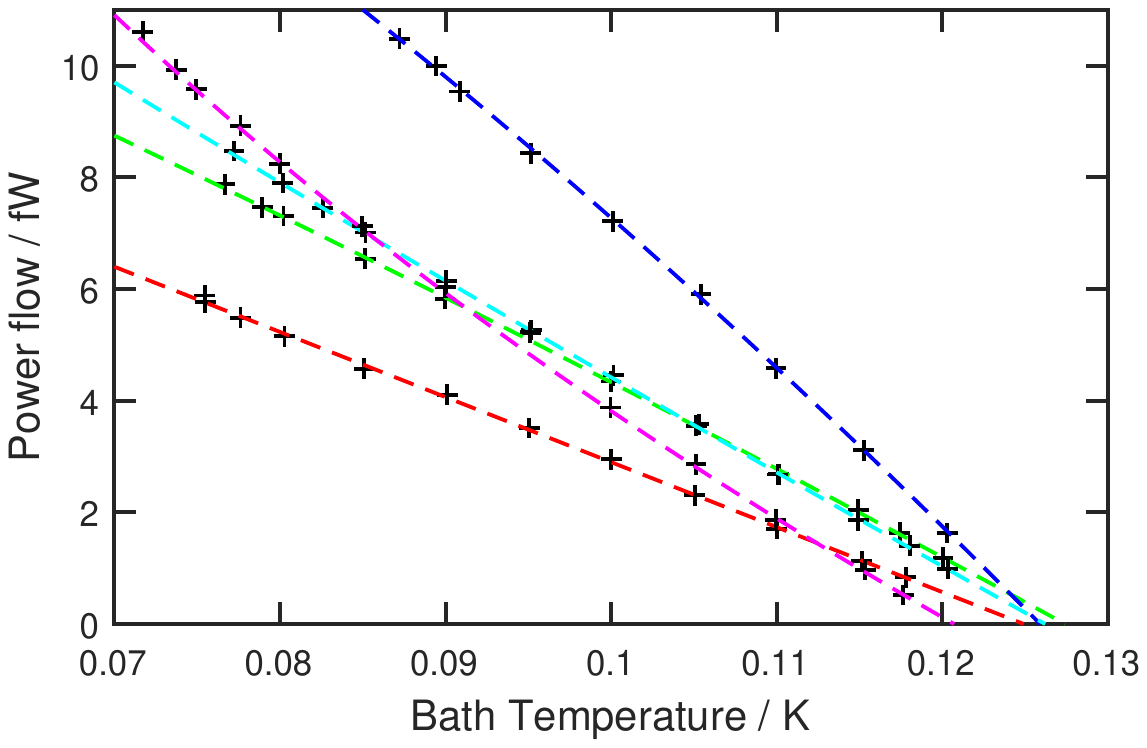}
\par\end{centering}
\caption{Measured power flow as a function of bath temperature. The different curves correspond to the devices listed in Table \ref{table1}: device 1 (red), 2 (green), 3 (blue), 4 (cyan), 5 (mauve).}
\label{figc}
\end{figure}

The purpose of the work described here was to observe the ballistic to diffusive transition in low-dimensional SiN$_{\rm x}$ bars, and therefore it is pertinent to comment on the structure and stoichiometry of the SiN$_{\rm x}$ used. The material was grown by LPCVD using dichlorosilane and ammonia. Because hydrogen and hydrochloric acid are both byproducts of the process, hydrogen can remain trapped, which affects physical characteristics. By increasing the flow of dichlorosilane, almost stress-free nitride can be formed through a silicon enriching process. The optical (633~nm) refractive index of the SiN$_{\rm x}$ used for our devices was typically in the range 2.0 to 2.3, corresponding to the near-stoichiometric limit, which is consistent with a measured tensile stress of 500~MPa. SiN$_{\rm x}$  of this composition is known to have many voids having scale sizes of around 10 nm. Our own surface roughness measurements using Atomic Force Microscopy (AFM) show structure having a log-normal height distribution covering the range 0.5-2.5~nm with a long tail out to 10~nm.

The primary experimental method requires the base temperature of the fridge to be varied whilst recording the power flowing onto the island of the device. A series of measurements were taken so that the experimental data could be corrected for voltage offsets and stray resistance. Figure~\ref{figB} shows a set of typical, calibrated IV curves, and Fig.~\ref{figc} shows the power flow as a function of bath temperature. Because of electrothermal feedback, the hot temperature $T_{h}$ essentially stays constant at the critical temperature $T_{c}$ of the bilayer during this process. The intercept on the abscissa is the critical temperature  of the bilayer. All of the $T_{c}$'s are within 7~mK of each other, with the suppression of device 5 being due to the bilayer having 6 normal-metal bars patterned on its surface \cite{ref30}. The power flow was then fitted to the functional form
\begin{equation}
P = K \left( T_{h}^{n} - T_{b}^{n} \right),
\end{equation}
where $K$ and $n$ are parameters, and $T_{b}$ was the temperature of the copper block that housed the chip. The thermal conductance $G$ can then be determined using $G = \partial G / \partial T_{h} = K n T_{h}^{n-1}$. Although $P$ is, strictly, measured with respect to variations in bath temperature, the symmetry of the expression allows $G$ to be calculated for variations in $T_{h}$. This method for measuring thermal conductance is standard in the TES community . Table 1 lists key parameters derived for the new devices measured; previous data has already been described \cite{ref9,ref13}.

\begin{figure}[h]
\noindent \begin{centering}
\includegraphics[trim = 1.0cm 1.0cm 7.0cm 21.0cm,width=8.0cm]{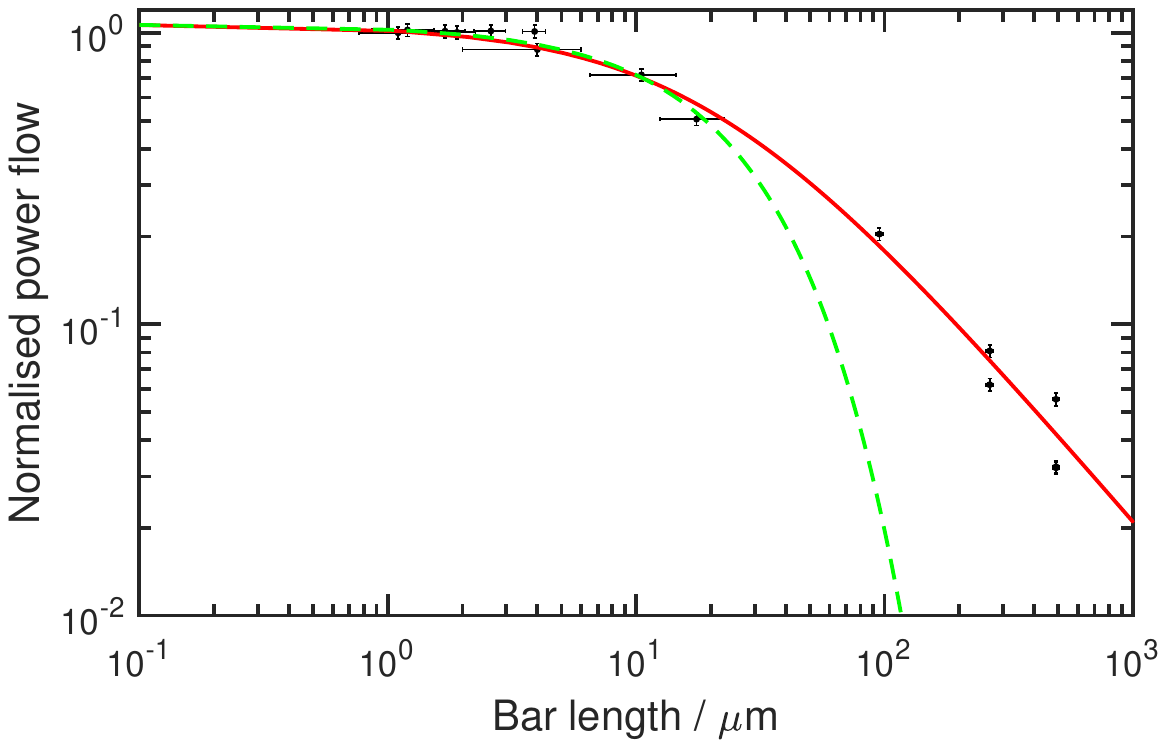}
\par\end{centering}
\caption{Normalised power flow as a function of leg length, taking into account the fact that each TES has 4 support legs. Solid red line: diffusive model having an attenuation length of 20~$\mu$m. Dashed green line: localised model having a localisation length of 25~$\mu$m.}
\label{figA}
\end{figure}

Fig.~\ref{figA} shows the normalised power flow, at the coldest bath temperature $80 \pm 5$~mK, as a function of leg length, where we have also included measurements of our earlier ballistic devices \cite{ref9}. Measurements on longer legs, which are entirely consistent with the results reported here, are not included because they add little information about the transition region. The powers plotted are normalised to the ballistic power expected on the basis of the effective number of elastic modes available, such that one would expect the limiting ballistic value to be unity. More specifically, normalisation was achieved as follows: For each each device, the width and height of the legs were used to calculate the dispersion curves of the modes . These calculations required finding full numerical solutions to the elastic wave equations \cite{ref9,ref13}. They were based on the bulk elastic constants (density, Young's modulus, and Poisson's ratio) of SiN$_{\rm x}$. Because of the long-wavelength nature of the phonons, this approach is suitable because it averages over any microstructure in the material. In any case, the dispersion curves are insensitive to the precise values of the bulk elastic constants used.

The cut-off frequencies of the modes were then used to calculate the ballistic power flow, $P_{\rm bal}$, between two heat baths; one held at the critical temperature of the bilayer and the other held at the lowest bath temperature used in the experiment:
\begin{equation}
P_{\rm bal} = \sum_{i} \int_{\nu_{i}}^{\infty} B(\nu,T_{h}) -  B(\nu,T_{b}) \, d \nu,
\end{equation}
where $\nu_{i}$ is the cut-off frequency of mode $i$, and
\begin{equation}
B(\nu,T) = \frac{h \nu}{ e^{h \nu / k T} -1}.
\end{equation}
Finally, the effective number of modes carrying heat was calculated through $N_{\rm eff} = P_{\rm bal}/P_{\rm qua}$, where
\begin{equation}
P_{\rm qua} = \int_{0}^{\infty} B(\nu,T_{h}) -  B(\nu,T_{b}) \, d \nu,
\end{equation}
is the ballistic power that would flow in a single mode that propagates at all frequencies. Experimentally, we found that typically $N_{\rm eff} = 5-7$ modes transported heat in our devices at the temperatures used, with $N_{\rm eff} = 4$ being the low-temperature limit. Finally, the normalised heat flow, shown in Fig.~\ref{figA}, was calculated through $\epsilon = P / 4P_{\rm qua} N_{\rm eff}$, where the factor of 4 accounts for each TES having 4 legs. It can be seen that all of the short-legged devices had a normalised power flow of near unity, confirming that the power flowing in ballistic legs can be calculated accurately using dispersion relationships based on bulk elastic constants, with no free parameters. The power measurement errors shown in Fig.~\ref{figA} correspond to $\pm$ 5~\%, which is conservative, and the errors in length, for the very short legs, arose because of the uncertainty in length associated with the gradual widening of the legs as they connected to their termination points. For ballistic legs this is not an issue, as the key effect of the constriction is to limit the modal throughput of the structure.

\section{Discussion}

It is sometimes said that at low temperatures, $<$~500~mK, anharmonic scattering process are not significant, and phonon scattering is caused by surface roughness. At 100 mK, however, thermal power is carried phonons having frequencies of less than 5 GHz, corresponding to wavelengths in SiN$_{\rm x}$  greater than 500 nm. A calculation of specularity, based on measured surface roughness data, as a function of phonon wavelength shows that the probability of a phonon being specularly reflected is essentially 100 \% for wavelengths of greater than 500 nm. Detailed numerical simulations also show that the reduction in power seen cannot be accounted for by surface scattering \cite{ref13}. A more appealing model is one based on density inhomogeneities in the amorphous SiN$_{\rm x}$. We have carried out extensive scattered-wave calculations of thermal transport, dividing few-mode dielectric bars into many thousands of elements \cite{ref13,ref16}. Each scattering section, described by a set of complex elastic-wave scattering parameters, was chosen to represent a density inhomogeneity of a few percent with a scale length of 5~nm, in correspondence with the distribution of known feature sizes in SiN$_{\rm x}$. In this case, localised transport is revealed, where the thermal conductance is reduced by the appearance of disordered resonant cells, spanning tens to hundreds of elements. In other words, travelling waves are reflected strongly by impedance discontinuities caused by the formation of Fabry-Perot-like resonators in the disorder of the material.

A feature of localised transport is that for bars that are long compared with the localisation length, the transmitted power $P$ varies exponentially with length
\begin{equation}
\label{eqn1}
P = P_{0} \exp[- L / L_{e}],
\end{equation}
where $P_{0}$ is the few-mode ballistic limit, and $L_{e}$ is a characteristic scale length. Another feature of localisation is that the conductances of long bars are predicted to vary widely from one physical sample to the next; although, this variation is reduced as the number of modes is increased, 2D and 3D transport is approached, and power is scattered laterally. In low-dimensional systems, the formation of high-Q resonant features exaggerates the variations in disorder from one sample to the next. Our own devices, and those of other groups, having legs that are hundreds of microns long, show measured conductance variations of typically $\pm$15 \%, but sometimes higher, even between notionally identical devices on the same wafer (verified by optical inspection and measurements of geometry and surface roughness). The variations seen in the longest legs in Fig.~\ref{figA} is a real effect, well above experimental error, which is tiny on the plot. Long, wide artificially surface-roughened, but otherwise identical crystalline Si bars show even more extreme variations \cite{ref17}, typically factors of 5, which is strongly indicative of localisation.

On Fig.~\ref{figA} we plot (\ref{eqn1}) for $L_{e} =$~25~$\mu$m, dashed green line, and it can be seen that an exponential dependence on length is ruled out.  It can also be seen that the short, ballistic legs have almost identical behaviour, leading to high levels of device uniformity, and that the sample-to-sample variations increase as bars are made longer. This trend has been seen in our work going back many years. Our simulations also show that attenuation lengths of less than 100~$\mu$m require RMS density inhomogeneities of over 20\%, which is possible, but seems high. In addition, the conductance repeatability of even long SiN$_{\rm x}$ bars is much better than our localised heat transport simulations would suggest, and seen in crystalline Si \cite{ref17}. Overall, we take these results to indicate that some level of localisation caused by density inhomogeneities is present in long bars, $>$~100~$\mu$m, but this does not explain why the conductance variations are quite small, and indeed why the power falls as $1/L$ in the case of long bars. The fact that sample-to-sample conductance variations in disordered SiN$_{\rm x}$ are substantially less than those seen in surface-roughened crystalline Si, and the complete inability to fit Fig.~\ref{figA} with an exponential, shows that phase-incoherent inelastic damping must be present in the amorphous material.

At the other extreme, inelastic scattering leads to fully diffusive transport. In this case, it is straightforward to show analytically that
\begin{equation}
\label{eqn2}
P = P_{0} (1 + L/L_{a})^{-1},
\end{equation}
where  $L_{a}$ is the amplitude attenuation length of the low-order elastic waves present: the travelling wave amplitude decays according to $a(z) = a(0) \exp \left( - z / L_{a} \right)$. (\ref{eqn2}) can be appreciated by differentiating each side with respect to temperature, and noting that $d P_{0} / d T = G_{\rm qua}$ is the quantum thermal conductance, giving
\begin{equation}
\label{eqn3}
\frac{1}{G} = \left( \frac{1}{G_{\rm qua}} + \frac{N}{G_{\rm qua}} \right)
\end{equation}
where $N = L/L_{a}$, the length of the bar in attenuation lengths. The interpretation is clear: the overall conductance comprises the quantum limit of conductance in series with $N$ cells, each of which behaves in a local sense ballistically so that it contributes an additional quantum-limited conductance.

Physically, the most likely scattering process corresponds to phase-incoherent absorption and reradiation by Two Level Systems (TLSs) in the amorphous material. TLSs are known to absorb ultrasonic waves at microwave frequencies, to lead to high specific heats, and to influence thermal conductance \cite{ref18,ref19,ref20,ref21,ref22}. We have measured the heat capacity of our SiN$_{\rm x}$ to be many hundreds of times higher than the Debye value, an effect that is traditionally attributed to TLSs \cite{ref23,ref24}. We also know that for very long legs, $>$ 400~$\mu$m, the conductances of our TESs fall as $1/L$.  It should be noted that although, traditionally, TLSs are used to describe non-Debye-like dissipative behaviour in disordered dielectrics, the precise physical origin is usually not known \cite{ref31}. In fact any non-harmonic dynamical behaviour that has non-equally spaced energy levels will lead to the saturation that TLSs are commonly used to represent. Here, we use the term TLS loosely to indicate the presence of any low-energy phase incoherent scattering process.

The red solid line in Fig.~\ref{figA} shows a typical diffusive model, having the form of (\ref{eqn2}), with $L_{a}=$ 20~$\mu$m. The data is consistent with diffusive transport. Notice that in both the cases of localised and diffusive transport, the best fits give normalised ballistic conductances that are slightly higher than unity, which can be easily accounted for by uncertainties in the bulk elastic constants of the material.

There is another peculiarity in our data that points to the role of TLSs. Short ballistic legs consistently have an $n$ of around 2.4, which is exactly what one would expect. Indeed the fully ballistic limit of the lowest order modes, which propagate at all frequencies, is $n=2$. As the temperature is increased and higher-order modes cut on, the value of $n$ increases accordingly. Our own very wide-leg devices, $>$~10~$\mu$m, and those of other groups, have $n$ in the range 3 to 4, which is characteristic of a highly-moded structure. The legs measured here have an $n$ of typically 1: this is reproducible, and something we have seen many times, going back several years, for long narrow legs. (Device 5 shows an anomalously low value of $n$, which is an artefact of the normal metal bars on the bilayer reducing the electrothermal feedback that holds the temperature of the TES constant as the temperature of the bath is varied, but this does not invalidate the flux measurement.)

As the length of a bar is increased through the ballistic to diffusive transition, keeping the number of propagating modes constant, the temperature dependence changes from $n=2.5$ to $n = 1$. Diffusion calculations based on travelling waves with TLS absorption do show that $n$ can fall well below 2 as a consequence of the elastic losses having a frequency and power dependence \cite{ref25}. Interestingly, the reason why this is not seen in wide-leg devices is that the increase in the number of modes with temperature, masks the effect on $n$ of the TLSs. Also, a value of $n=$~0.5 can be seen in the data of Schwab \cite{ref1}, up to a temperature of 200~mK, at which point the gradient changes to $n=$~2.7, which is characteristic of the temperature range over which the frequency and power dependence of TLSs would be influential. Our measured elastic-wave attenuation length of 20~$\mu$m is slightly low, but generally in good agreement with ultrasonic, mechanical and thermal measurements of phonon mean free path made on bulk amorphous dielectrics \cite{ref20,ref22,ref23,ref26}. It is also interesting to estimate the Q-factor relating to `internal friction'. Using $Q^{-1} = \lambda / \pi L_{a}$, where $\lambda \approx$~500~nm is an approximate characteristic thermal wavelength, we find $Q^{-1} =$~8$\times$10$^{-3}$, which again is slightly high, but comparable with measured Q$^{-1}$ factors of amorphous bulk materials. Typically in the range 10$^{-3}$ to 10$^{-4}$ \cite{ref21,ref27}. It has been noted by others that the vibrational modes of mesoscopic systems have anomalously low Q values compared with larger systems fabricated from the same material \cite{ref4}.

We make two final comments. The first is that we are not entirely sure about the degree to which a very thin layer of SiO$_{2}$, used as an etch stop, is removed from the underside of the legs during processing. Zink and Hellman \cite{ref23} noticed that the nature of the residual underlayer can affect thermal properties at low temperatures. The scattering processes described here nevertheless prevail in the case where surface contamination is present, say if patches of SiO$_{2}$ remain on one side of the legs after processing; it is simply that there is an additional contribution to the disorder, and perhaps surface states. The second  is that Fig.~\ref{figA} shows that even when the leg length is shorter than 1~$\mu$m, there is no evidence of a rapid increase in thermal flux above the ballistic travelling-wave limit, indicating that there is no measurable evanescent coupling \cite{ref28,ref29}.

\section{Conclusion}

We have observed directly the diffusive to ballistic transition in heat flow along low-dimensional SiN$_{\rm x}$ bars. The nature of the transition is indicative of a diffusive process, but the systematic increase in device-to-device variation as the bars are made longer is strongly indicative of localisation: Fabry-Perot resonances caused by disorder in the material. The thermal elastic-wave attenuation length has been determined to be 20~$\mu$m, which to our knowledge is the first measurement of this important parameter. An anomalous value of $n \approx$~1 in the case of long, narrow, few-mode bars, points to the role of TLSs. Overall, we favour a model where the dissipative losses associated with TLSs bring about diffusive transport and also dampen the Q factors of the localised resonant cavities formed by disorder.

Ballistic legs, $L<$~10~$\mu$m, give highly uniform device-to-device behaviour because of the elimination of localisation, but for some applications, the thermal conductances achieved are not low enough. Given that the characteristic wavelength of thermal phonons is approximately 1~$\mu$m at these temperatures, and given that the travelling-wave attenuation length has been measured to be 20~$\mu$m, it is fully realistic to micro-engineer patterned phononic filters, Mach-Zehnder interferometers, that use interference effects to control the flow of heat. If diffusive transport is needed, a longer leg must be used, but some uniformity in behaviour will be lost. The observed coherence length, and the nature of the scattering mechanism present, will have a strong influence on the thermal fluctuation noise in the legs. In the case of an inelastic process, thermal exchange noise can take place between the losses within a coherence length of the end of a bar and the bath. We are keen to measure fluctuation noise as a function of leg length as the functional form would shed further light on the scattering processes present.

\begin{acknowledgments}

Emily Williams gratefully acknowledges support from NanoDTC EPSRC Grant EP/L015978/1 during the course of this work.

\end{acknowledgments}

\end{document}